\documentclass{isprs} 
\usepackage{subfigure}
\usepackage{setspace}
\usepackage{geometry} 
\usepackage{epstopdf}
\usepackage[labelsep=period]{caption}  
\usepackage[british]{babel} 
\usepackage[hang]{footmisc}

\usepackage{caption}
\usepackage{amsmath}
\usepackage{amssymb}
\usepackage{todonotes}
\usepackage{graphicx}
\usepackage{algorithm}
\usepackage{algpseudocode}
\usepackage{subfigure}
\usepackage{booktabs}
\usepackage{multirow}
\usepackage{xcolor}
\definecolor{DarkGreen}{RGB}{0,128,0}
\usepackage{colortbl}
\usepackage{tikz}        
\usepackage{pgfplots}
\usepgfplotslibrary{colorbrewer}
\usepackage{url}

\usepackage{placeins}
\usepackage[authoryear]{natbib}
\usepackage[acronym]{glossaries}
\makeglossaries
\newacronym{3dgs}{3DGS}{3D Gaussian Splatting}
\newacronym{sfm}{SfM}{Structure-from-Motion}
\newacronym{nerf}{NeRF}{Neural Radiance Fields}
\newacronym{ssim}{SSIM}{structural similarity}
\newacronym{nvs}{NVS}{novel view synthesis}
\newacronym{mvs}{MVS}{Multi-view Stereo}
\newacronym{tnt}{TnT}{Tanks and Temples}

\begin{document}

\title{CDGS: Confidence-Aware Depth Regularization for 3D Gaussian Splatting}
\date{}

\author{
 Qilin Zhang\textsuperscript{1}, Olaf Wysocki\textsuperscript{1}, Steffen Urban,  Boris Jutzi\textsuperscript{1}}

\address{
	\textsuperscript{1}Photogrammetry and Remote Sensing, TUM School of Engineering and Design, Technical University of Munich (TUM), \\Munich, Germany - (qilin.zhang, olaf.wysocki, boris.jutzi)@tum.de\\
}

\commission{Technical Commission II}{}

\abstract{
\gls{3dgs} has shown significant advantages in \gls{nvs}, particularly in achieving high rendering speeds and high-quality results. 
However, its geometric accuracy in 3D reconstruction remains limited due to the lack of explicit geometric constraints during optimization. 
This paper introduces CDGS, a confidence-aware depth regularization approach developed to enhance \gls{3dgs}. 
We leverage multi-cue confidence map of monocular depth estimation and sparse \gls{sfm} depth to adaptively adjusts depth supervision during the optimization process. 
Our method demonstrates improved geometric detail preservation in early training stages and achieves competitive performance in both \gls{nvs} quality and geometric accuracy.
Experiments on the public available Tanks and Temples benchmark dataset show that our method achieves more stable convergence behavior and more accurate geometric reconstruction results, with improvements of up to 2.31 dB in PSNR for \gls{nvs} and consistently lower geometric errors in M3C2 distance metrics. Notably, our method reaches comparable F-scores to the original \gls{3dgs} with only 50\% of the training iterations. 
We expect this work will facilitate the development of efficient and accurate 3D reconstruction systems for real-world applications such as digital twin creation, heritage preservation, or forestry applications. 
}


\maketitle
\footnotetext[1]{\url{https://github.com/zqlin0521/cdgs-release}} 
\section{Introduction}\label{Introduction}
\glsresetall
The development of \gls{nvs} methods, particularly with the advent of \gls{nerf}, has brought revolutionary changes to the field of 3D reconstruction. 
\gls{nerf} sets a benchmark in photorealistic rendering by capturing detailed scene features \citep{mildenhall2021nerf}. 
However, \gls{3dgs} \citep{kerbl20233d} has emerged as an effective alternative to \gls{nerf} and achieves a better balance between rendering efficiency and reconstruction quality. 
Unlike \gls{nerf}, which models scenes using dense neural representations to capture complex lighting and shading effects, \gls{3dgs} represents scenes with anisotropic 3D Gaussian primitives. 
By optimizing the positions, orientations, and appearances of these primitives based on input image data and using an efficient tile-based rasterization technique, 3DGS achieves real-time rendering while maintaining high visual quality. 

Despite its strengths, \gls{3dgs} struggles with accurately reconstructing 3D structures due to multi-view inconsistencies inherent to 3D Gaussian primitives \citep{huang20242d}. 
Integrating additional geometric cues, such as depth information, offers a promising solution to these challenges \citep{chung2024depth}. 
However, monocular depth estimation models can lack robustness under diverse scene conditions, potentially compromising the consistency and reliability of depth-based optimization. 
Furthermore, although geometric evaluation is essential for assessing the accuracy and reliability of 3D reconstructions \citep{petrovska2024vision}, most existing studies prioritize enhancing the 2D rendering quality of \gls{3dgs}. 
The impact of depth information integration on the 3D reconstruction accuracy of \gls{3dgs} has received limited systematic analysis. 
This gap underscores the need for both a more stable depth-based optimization approach and a comprehensive evaluation framework for 3D reconstruction tasks in \gls{3dgs}.

\begin{figure*}[ht!]
    \centering
    \includegraphics[width=\textwidth]{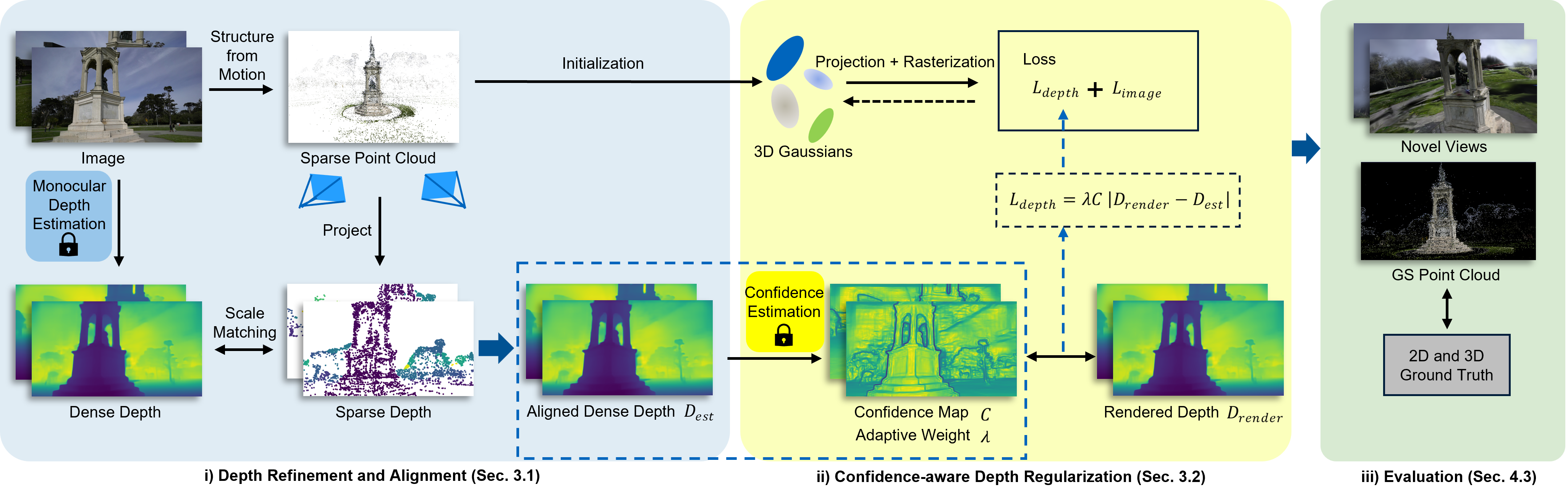}
    \caption{Overview of our confidence-aware depth regularization framework for \gls{3dgs}. Our method introduces three key components: i) \textit{depth refinement and alignment}, ii) \textit{confidence-aware depth regularization} through multi-cue feature analysis, and \\iii) comprehensive \textit{2D and 3D evaluation} metrics for assessing both rendering quality and geometric accuracy. This framework enables stable optimization and improved reconstruction results.}
    \label{fig:overview}
\end{figure*}

\begin{sloppypar}
Building on these insights, we introduce CDGS, a confidence-aware depth-based optimization strategy for \gls{3dgs}. 
As shown in Figure \ref{fig:overview}, CDGS enhances the original \gls{3dgs} method through two key components: i) \textit{depth refinement and alignment} and ii) \textit{confidence-aware depth regularization}. 
Our approach first utilizes the recently developed Depth Anything V2 model \citep{yang2024depth} to obtain initial depth maps from input images. 
These depth maps are further aligned with sparse depth data from \gls{sfm} using a gradient descent method. 
For each aligned depth map, we generate a confidence map by analyzing features from both the depth map and its corresponding RGB image. 

During optimization, while the original \gls{3dgs} combines pixel-wise L1 loss (absolute differences between rendered and ground truth images) with \gls{ssim} loss, our method introduces an additional depth regularization term. 
This regularization incorporates both depth normalization and dynamic loss adjustment based on the estimated confidence maps, thereby improving 3D reconstruction accuracy. 
We comprehensively evaluate our method from both 2D and 3D perspectives and analyze the performance throughout the optimization process.

In summary, the main contributions of this work are as follows:

\begin{itemize}
    \item We propose a \textit{depth refinement and alignment} method that aligns monocular depth estimates with sparse depth data from \gls{sfm} through gradient descent optimization, improving geometric consistency across multiple views.
    
    \item We introduce a \textit{confidence-aware depth regularization} term that generates confidence maps by analyzing features from both depth maps and RGB images, and adaptively adjusts depth loss weights to achieve stable optimization.
    
    \item We present comprehensive \textit{2D and 3D evaluations} on benchmark datasets, providing systematic comparisons with state-of-the-art \gls{3dgs} variants and detailed analysis of optimization behavior in both \gls{nvs} and geometric reconstruction.

\end{itemize}
\end{sloppypar}
\section{Related Work}\label{Related Work}
\begin{sloppypar}
In this section, we first introduce the development of \gls{nvs} techniques from traditional geometric approaches to recent neural rendering methods. 
We then discuss the integration of depth information in \gls{3dgs} and various depth-based optimization strategies. 
Finally, we describe the fundamental formulation of \gls{3dgs} that serves as the basis for our method. 

\textbf{Novel View Synthesis.} 
\gls{nvs} aims to generate novel views of a scene from unseen perspectives. It commonly relies on 3D reconstruction techniques to capture spatial structures and preserve visual details. 
A fundamental step in \gls{nvs} involves estimating the camera’s intrinsic and extrinsic parameters \citep{mueller2019image}, a task typically performed using Structure-from-Motion (SfM) methods \citep{ullman1979interpretation}. 
These methods leverage feature matching across multiple views and epipolar geometry to recover both camera poses and sparse 3D scene structure. 
Building upon \gls{sfm} results, \gls{mvs} techniques \citep{tomasi1992shape} can further enhance the reconstruction by creating denser 3D models. 

Beyond traditional methods such as \gls{sfm} and \gls{mvs}, deep learning has brought new advances to \gls{nvs}, particularly with \gls{nerf}. 
\gls{nerf} represents scenes as continuous volumetric radiance fields encoded by a neural network \citep{mildenhall2021nerf}. It achieves photorealistic rendering by modeling the color and density of each 3D point based on the viewing direction. 
Despite its success in generating high-quality novel views, NeRF’s computational demands often limit its efficiency in real-time applications. 
To meet the demand for faster rendering, \gls{3dgs} has emerged as a promising alternative. It offers an efficient approach by using \(\alpha\)-blending rasterization rather than the computationally intensive volume rendering. 
By optimizing the spatial distribution, scales, rotations, and opacities of the Gaussian primitives, \gls{3dgs} achieves both real-time rendering and high-quality reconstruction \citep{chung2024depth}. 

\textbf{Gaussian Splatting with Depth Information.} 
Recent works have explored integrating monocular depth estimation into \gls{3dgs} optimization. 
Various depth estimation models have been employed, such as DPT \citep{ranftl2021vision}, which has been applied in \citet{chung2024depth} and \citet{turkulainen2024dn}, and ZoeDepth \citep{bhat2023zoedepth}, which has been utilized in \citet{li2024dngaussian} and \citet{zhu2023fsgs}. 
Recent advances in monocular depth estimation have led to new models such as Depth Anything V2 \citep{yang2024depth}. 
In this work, we investigate the integration of this recent depth estimation model into \gls{3dgs} optimization.

Many existing methods adopt the depth-based regularization framework introduced by \citet{chung2024depth}, which extends the original \gls{3dgs} rasterization pipeline to produce rendered depth maps for geometric supervision. 
Expanding on this framework, researchers have proposed strategies to enhance \gls{3dgs} performance, such as achieving comparable \gls{nvs} quality with fewer training images \citep{zhu2023fsgs} and improving rendering details \citep{li2024dngaussian}. 
In terms of depth regularization, different approaches have been explored, including fixed-weight balancing between depth and image losses \citep{liu2024endogaussian}, segmented depth regularization with varying emphases \citep{li2024dngaussian}, and early stopping strategies based on depth loss variations \citep{chung2024depth}.
However, these approaches rely on the estimated depth values without explicitly considering their reliability, which may lead to inconsistent optimization behavior across different scenes. 
Furthermore, while existing methods have demonstrated improvements in 2D image synthesis quality, there has been limited systematic analysis of how depth information affects the 3D geometric accuracy.  
To tackle these limitations, our method proposes a confidence-aware depth regularization strategy that selectively incorporates depth supervision based on multi-cue reliability assessment, aiming to achieve more stable optimization in both novel view synthesis and geometric reconstruction tasks.
\end{sloppypar}

\textbf{3D Gaussian Splatting Formulation.}
\gls{3dgs} represents 3D scenes using a collection of 3D Gaussian primitives. Each primitive is parameterized by its center \( \mu \in \mathbb{R}^3 \), scale \( s \in \mathbb{R}^3 \), rotation (quaternion) \( q \in \mathbb{R}^4 \), opacity \( \alpha \in \mathbb{R} \), and color features \( f \in \mathbb{R}^K \). The complete parameter set for the \( i \)-th Gaussian is denoted as \( \theta_i = \{ \mu_i, s_i, q_i, \alpha_i, f_i \} \) \citep{kerbl20233d}. The Gaussian function is expressed as:
\begin{equation}
G_i(x) = \exp \left(-\frac{1}{2} (x - \mu_i)^T \Sigma_i^{-1} (x - \mu_i)\right),
\end{equation}
where the covariance matrix \( \Sigma \) is determined by scale \( s \) and rotation \( q \).

For rendering, \gls{3dgs} employs an efficient rasterization pipeline that performs \(\alpha\)-blending of projected Gaussians. The color \( C \) of each pixel is computed by blending contributions from \( N \) overlapping Gaussians:
\begin{equation}
C = \sum_{j \in N} c_j \alpha_j T_j,
\end{equation}
where \( T_j = \prod_{k=1}^{j-1} (1 - \alpha_k) \) represents the accumulated transparency. Here, \( c_j \) and \( \alpha_j \) denote the color and opacity of the \( j \)-th Gaussian, respectively. This formulation naturally handles occlusion by giving priority to Gaussians closer to the camera.

The depth value \( D \) for each pixel is computed through normalized \(\alpha\)-weighted averaging:
\begin{equation}
D = \frac{\sum_{j \in N} d_j \alpha_j T_j}{\sum_{j \in N} \alpha_j T_j},
\end{equation}
where \( d_j = (R_i p_j + T_i)_z \) represents the depth of the \( j \)-th Gaussian relative to the \( i \)-th camera. This normalization ensures robust depth estimation even in regions with sparse Gaussian coverage \citep{chung2024depth}.
\section{Methodology}\label{Methodology}
\begin{sloppypar}
In this section, we present our methodology for enhancing \gls{3dgs} through two key components: i) \textit{depth refinement and alignment}, and ii) \textit{confidence-aware depth regularization}. 
The proposed method integrates depth information into the optimization process by considering its reliability and adaptively adjusting its influence during training. 
\subsection{Depth Refinement and Alignment}
To obtain reliable depth information for regularization, we propose a two-stage process: initial depth estimation and subsequent depth refinement through alignment with sparse geometric constraints.

Given an input image $I \in \mathbb{R}^{H \times W \times 3}$, we first obtain an initial depth map $D_{init} \in \mathbb{R}^{H \times W}$ using a pre-trained monocular depth estimation model $f$:
\begin{equation}
D_{init} = f(I)
\end{equation}
where $f$ represents depth estimation models, for example, ZoeDepth~\citep{bhat2023zoedepth} or Depth Anything V2~\citep{yang2024depth}. 

To provide geometric constraints for depth refinement, we generate a target depth map $D_{target}$ by projecting the sparse point cloud obtained from \gls{sfm}. For each 3D point $P \in \mathbb{R}^3$, its projected depth value at pixel coordinate $p$ is computed as:
\begin{equation}
D_{target}(p) = [K(RP + T)]_z
\end{equation}
where $K \in \mathbb{R}^{3 \times 3}$ is the camera intrinsic matrix, $R \in \mathbb{R}^{3 \times 3}$ and $T \in \mathbb{R}^3$ are the camera extrinsic parameters, and $[\cdot]_z$ denotes the z-component of the projected point.

Following \citet{chung2024depth}, we refine the initial depth estimates by aligning them with the sparse point cloud data through an optimization process. The objective function is formulated as:
\begin{equation}
\min_{\alpha, \beta} \sum_{p \in \Omega} w(p)(D_{target}(p) - (\alpha D_{init}(p) + \beta))^2 + \lambda R(\alpha D_{init} + \beta)
\end{equation}
where:
\begin{itemize}
    \item $\Omega$ denotes the set of pixels with valid point cloud projections
    \item $w(p)$ weights each pixel based on its point cloud reprojection error
    \item $\alpha$ and $\beta$ are scale and shift parameters for depth alignment
    \item $\lambda$ balances the alignment term and regularization term
\end{itemize}

The regularization term $R(\cdot)$ ensures physical validity by penalizing negative depth values:
\begin{equation}
R(D) = \sum_{p} \max(0, -D(p))^2
\end{equation}
We solve this optimization problem using gradient descent to obtain the optimal scale and shift parameters. 
During the optimization, we compute the alignment loss $l_a$ as the weighted mean squared error between the transformed and target depth:
\begin{equation}
l_a = \frac{1}{|\Omega|}\sum_{p \in \Omega} w(p)(D_{target}(p) - D_{trans}(p))^2
\end{equation}
where $D_{trans}(p) = \alpha D_{init}(p) + \beta$ represents the transformed initial depth using the current scale and shift parameters. 
This alignment loss not only guides the optimization process but also serves as an important indicator of depth estimation reliability, which is utilized in our subsequent confidence-aware depth regularization. 

The refined depth map provides geometric constraints for the following optimization process, while the alignment loss $l_a$ offers valuable information about the overall quality of depth estimation and alignment, enabling adaptive adjustment of depth supervision in the next stage.

\subsection{Confidence-aware Depth Regularization}
Building upon the refined depth maps and alignment quality information obtained from the previous stage, we propose a confidence-aware regularization scheme that adaptively incorporates depth supervision into the \gls{3dgs} optimization process. Inspired by \citet{wysocki2023scan2lod3}, our approach combines local depth reliability assessment with global alignment quality to guide the optimization.

We evaluate the reliability of depth values through a confidence map $C$ that integrates multiple complementary features:
\begin{equation}
C = w_e C_e + w_t C_t + w_g C_g
\end{equation}
where:
\begin{itemize}
\item $C_e$ captures edge-aware confidence using Canny edge detection:
\begin{equation}
C_e = 1 - \frac{E(I)}{255}
\end{equation}
where $E(I)$ represents the edge map of image $I$

\item $C_t$ measures texture reliability using the Laplacian operator:
\begin{equation}
   C_t = 1 - \frac{|\nabla^2 I|}{\max(|\nabla^2 I|)}
\end{equation}

\item $C_g$ evaluates depth consistency through spatial gradients:
\begin{equation}
   C_g = \frac{1}{|\nabla D| + \epsilon}
\end{equation}
where $\nabla D$ represents the spatial gradients of the refined depth map, and $\epsilon$ is a small constant to prevent numerical instability
\end{itemize}
The weights are chosen to emphasize texture reliability while maintaining balanced contributions from edge and gradient features, with $w_t > w_g > w_e$.

Based on the confidence map, we formulate the depth supervision term as:
\begin{equation}
   \mathcal{L}_{depth} = \frac{1}{|\Omega|}\sum_{p \in \Omega} C(p)|D_{render}(p) - D_{est}(p)|
\end{equation}
where $\Omega$ denotes the set of valid depth pixels, $D_{render}(p)$ is the rendered depth from the current \gls{3dgs} model, and $D_{est}(p)$ is the refined depth value from our alignment stage. 
The confidence map $C(p)$ ensures that the supervision focuses on reliable depth regions.

To adaptively balance the depth supervision, we introduce a weight term $\lambda_d(l_a)$ based on the global alignment quality:
\begin{equation}
\lambda_d(l_a) = \lambda_{max}e^{-kl_a}
\end{equation}
where $\lambda_{max}$ controls the maximum influence of depth supervision, and $k$ determines how quickly the weight decreases with increasing alignment loss $l_a$. This exponential form ensures that depth supervision is stronger when the alignment quality is better (lower $l_a$).

For image reconstruction, we follow the original \gls{3dgs} formulation:
\begin{equation}
\begin{split}
   \mathcal{L}_{image} = & (1-\lambda_{dssim})\|I_{pred} - I_{gt}\|_1 \\ 
   & + \lambda_{dssim}(1-SSIM(I_{pred}, I_{gt}))
\end{split}
\end{equation}
where $\lambda_{dssim}$ balances the contribution of L1 loss and structural similarity.

Our final optimization objective combines image reconstruction with confidence-aware depth supervision:
\begin{equation}
\mathcal{L} = \mathcal{L}_{image} + \lambda_d(l_a)\mathcal{L}_{depth}
\end{equation}

This confidence-aware regularization scheme effectively integrates depth information by considering both pixel-wise depth reliability and global alignment quality. 
The local confidence weighting helps focus the supervision on reliable regions, while the adaptive global weight automatically adjusts the strength of depth supervision based on the overall alignment quality, and leads to more stable optimization behavior. 
\end{sloppypar}
\section{Experiments}\label{Experiments}
\begin{sloppypar}
We conducted comprehensive experiments to evaluate the effectiveness of our proposed CDGS method. 
Our evaluation framework encompassed both novel view synthesis quality and geometric reconstruction accuracy through extensive comparisons with state-of-the-art methods. 
This section first describes the experimental setup, including datasets and implementation details, followed by detailed qualitative and quantitative analyses. 
\subsection{Datasets}
We evaluated our method on the \gls{tnt} benchmark dataset \citep{knapitsch2017tanks}, which provides multi-view images and high-precision laser-scanned ground truth geometry. 
The selected scenes included complex outdoor environments (Barn, Truck, Caterpillar, Ignatius) and detailed indoor scenarios (Meeting Room, Church), which featured diverse geometric complexities and lighting conditions \citep{urban2017lafida}. 
Following the protocol established by \citet{chung2024depth}, we employed a random sampling strategy where each captured view was assigned to either the training or testing set with equal probability. 
This probabilistic split strategy ensured a balanced distribution between training and testing data while maintaining well-distributed testing views across different viewpoints. 

Our evaluation framework incorporated both 2D and 3D metrics to provide a comprehensive assessment of the method's performance. 
For 2D novel view synthesis quality, we utilized three complementary metrics: PSNR for pixel-wise reconstruction accuracy, SSIM for structural similarity assessment, and LPIPS for perceptual quality evaluation.
The 3D geometric accuracy assessment combined the official evaluation protocol of Tanks and Temples \citep{knapitsch2017tanks} with the M3C2 distance metric \citep{lague2013accurate}.  
First, following the official evaluation protocol of Tanks and Temples, we performed a systematic point cloud registration process between the reconstructed geometry and ground truth, followed by the computation of precision and recall metrics. 
The protocol included F-score calculations at multiple distance thresholds to quantify the reconstruction quality across different levels of geometric detail. 
Additionally, inspired by \citet{jager2024hologs}, we utilized the M3C2 distance metric to provide a more detailed analysis of the geometric accuracy. 
This metric enabled a precise measurement of the local distances between the reconstructed and ground truth surfaces, offering insights into the fine-scale geometric fidelity of our reconstruction results. 

\begin{figure*}[ht!]
    \centering
    \begin{minipage}[t]{0.24\textwidth}
        \centering
        \includegraphics[width=\textwidth]{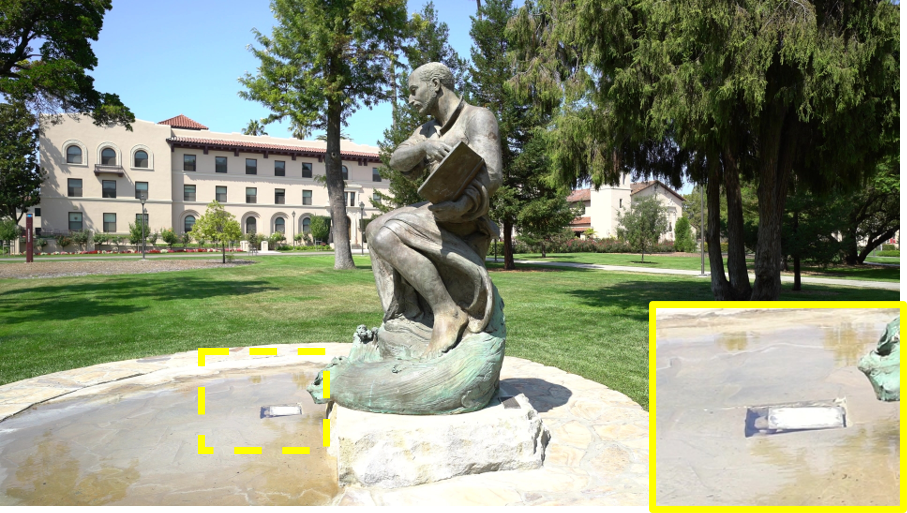}
        \subfigure{}
    \end{minipage}%
    \begin{minipage}[t]{0.24\textwidth}
        \centering
        \includegraphics[width=\textwidth]{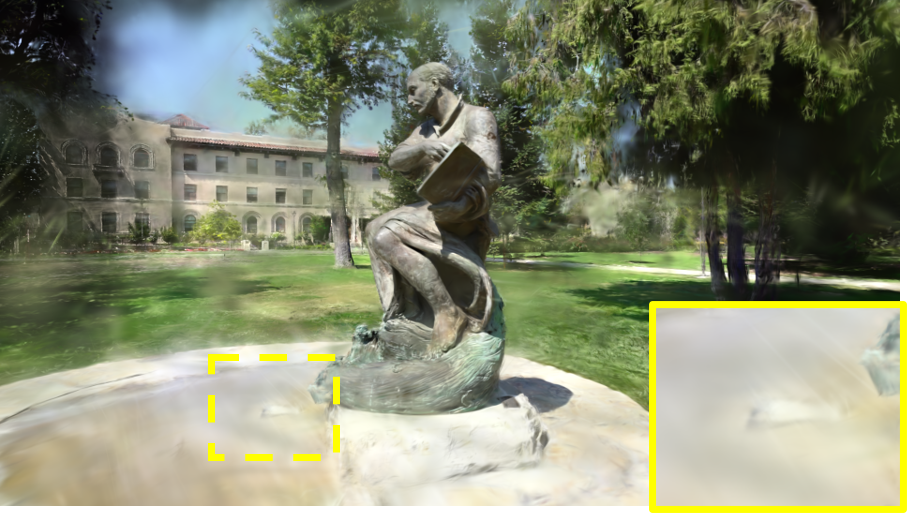}
        \subfigure{}
    \end{minipage}%
    \begin{minipage}[t]{0.24\textwidth}
        \centering
        \includegraphics[width=\textwidth]{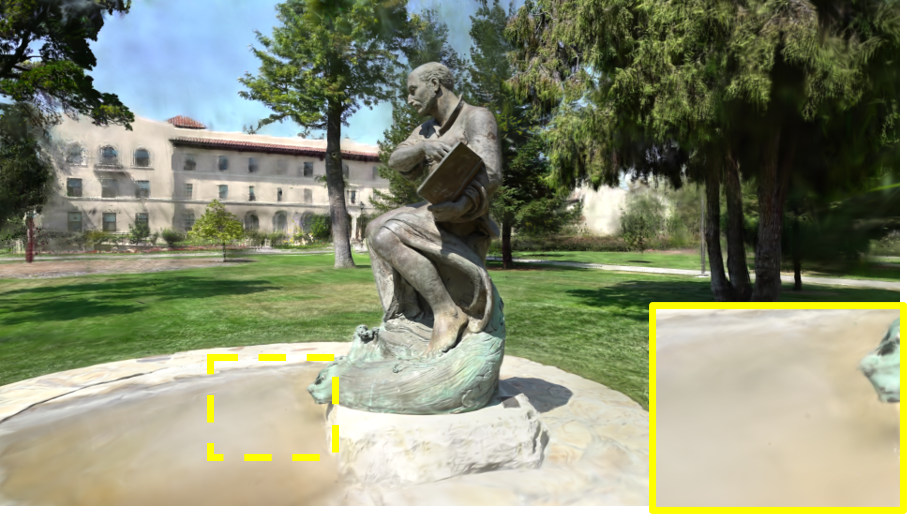}
        \subfigure{}
    \end{minipage}%
    \begin{minipage}[t]{0.24\textwidth}
        \centering
        \includegraphics[width=\textwidth]{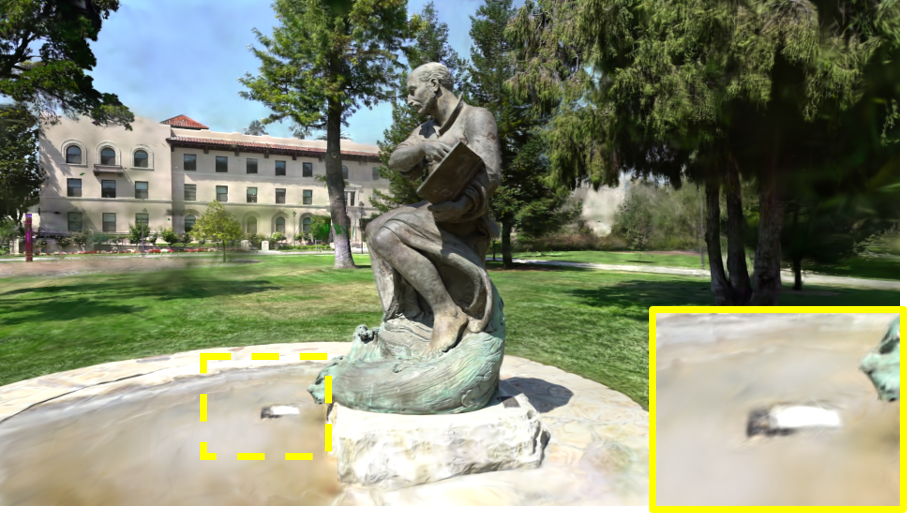}
        \subfigure{}
    \end{minipage}

    \begin{minipage}[t]{0.24\textwidth}
        \centering
        \includegraphics[width=\textwidth]{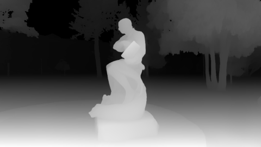}
        \subfigure{}
    \end{minipage}%
    \begin{minipage}[t]{0.24\textwidth}
        \centering
        \includegraphics[width=\textwidth]{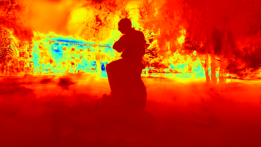}
        \subfigure{}
    \end{minipage}%
    \begin{minipage}[t]{0.24\textwidth}
        \centering
        \includegraphics[width=\textwidth]{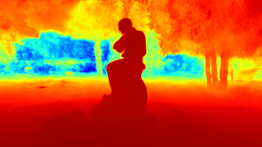}
        \subfigure{}
    \end{minipage}%
    \begin{minipage}[t]{0.24\textwidth}
        \centering
        \includegraphics[width=\textwidth]{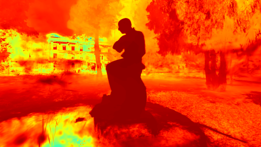}
        \subfigure{}
    \end{minipage}

    \begin{minipage}[t]{0.24\textwidth}
        \centering
        \includegraphics[width=\textwidth]{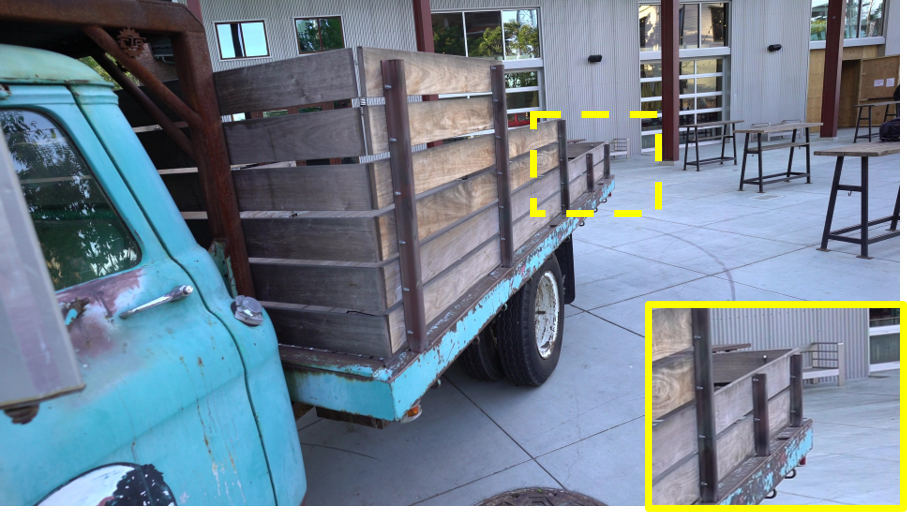}
        \subfigure{}
    \end{minipage}%
    \begin{minipage}[t]{0.24\textwidth}
        \centering
        \includegraphics[width=\textwidth]{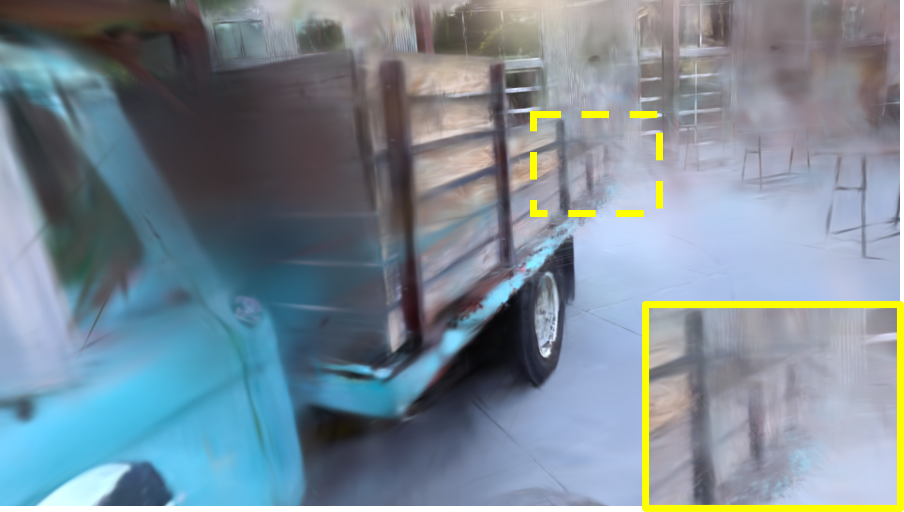}
        \subfigure{}
    \end{minipage}%
    \begin{minipage}[t]{0.24\textwidth}
        \centering
        \includegraphics[width=\textwidth]{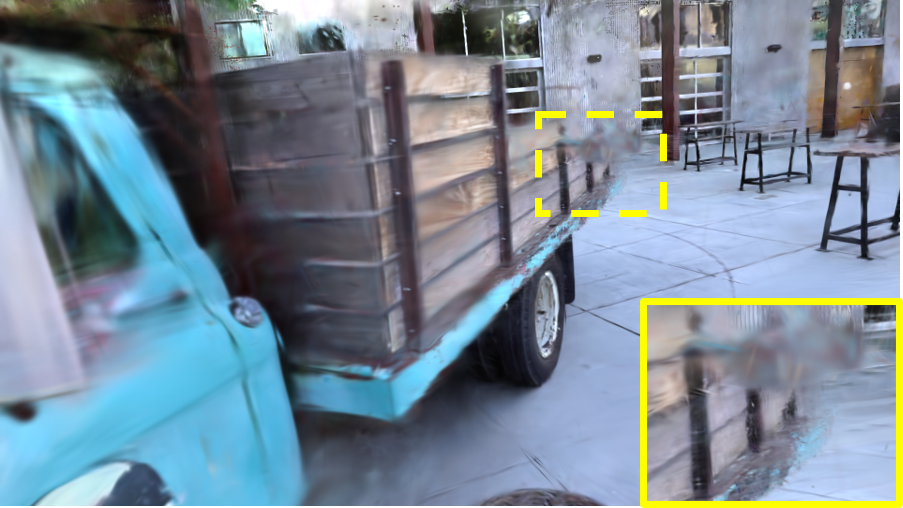}
        \subfigure{}
    \end{minipage}%
    \begin{minipage}[t]{0.24\textwidth}
        \centering
        \includegraphics[width=\textwidth]{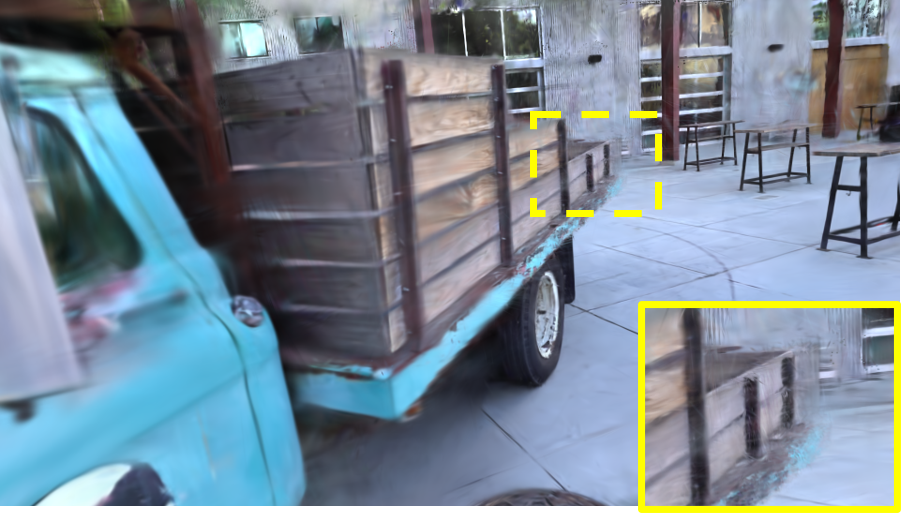}
        \subfigure{}
    \end{minipage}

    \begin{minipage}[t]{0.24\textwidth}
        \centering
        \includegraphics[width=\textwidth]{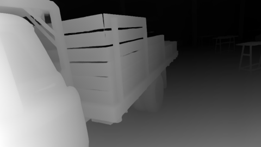}
        \subfigure{Ground Truth}
    \end{minipage}%
    \begin{minipage}[t]{0.24\textwidth}
        \centering
        \includegraphics[width=\textwidth]{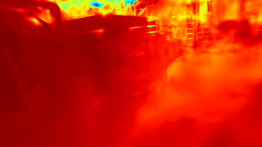}
        \subfigure{3DGS}
    \end{minipage}%
    \begin{minipage}[t]{0.24\textwidth}
        \centering
        \includegraphics[width=\textwidth]{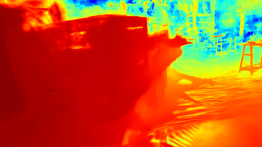}
        \subfigure{DRGS}
    \end{minipage}%
    \begin{minipage}[t]{0.24\textwidth}
        \centering
        \includegraphics[width=\textwidth]{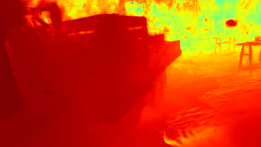}
        \subfigure{CDGS (Ours)}
    \end{minipage}
    \caption{Qualitative comparison of \gls{nvs} results on the Ignatius (top two rows) and Truck (bottom two rows) scenes at iteration 9,000. For each scene: rows 1\&3 show synthesized RGB images, rows 2\&4 present corresponding depth maps. Reference depth maps are generated using Depth Anything V2, and comparison depth maps are rendered from respective 3D representations. Applying our method required an additional preprocessing time of 1.5 seconds per image on average, ensuring its uniform applicability across all inputs. Yellow boxes highlight regions where our method better preserves geometric and radiometric details.}
    \label{fig:ignatius_comparison}
\end{figure*}

\subsection{Implementation details}
Our implementation was built upon the official codebase of \gls{3dgs} \citep{kerbl20233d}, and incorporated the depth-aware rasterization pipeline from DRGS \citep{chung2024depth}. 
We implemented our method using PyTorch and trained all models on a single NVIDIA Tesla V100-PCIE-32GB GPU. 
For depth estimation, we primarily used Depth Anything V2 \citep{yang2024depth}, while also supporting ZoeDepth \citep{bhat2023zoedepth} as an alternative monocular depth estimator. 
Following \citet{chung2024depth}, for depth alignment optimization, we employed Adam optimizer with an initial learning rate of 1.0 and an exponential decay schedule, with settings of outlier pruning ratio ($1\times10^{-3}$) and convergence threshold ($1\times10^{-5}$). 
For all experiments, we maintained the original optimization parameters and training pipeline from \citet{kerbl20233d} and trained for 30,000 iterations for fair comparison. 
For baseline comparison, we disabled the early stopping mechanism in DRGS \citep{chung2024depth} which typically terminates training before 2,000 iterations, to ensure a fair evaluation of its full potential. 
The key parameters specific to our method were the confidence map weights ($w_t=0.5$, $w_g=0.3$, and $w_e=0.2$) and depth loss adaptation parameters ($\lambda_{max}=0.6$ and $k=150$). 
The depth alignment process and the generation of confidence maps took approximately 1.5 seconds per image. 
The whole training process took approximately 2 hours per scene on our hardware configuration.
\end{sloppypar}
\subsection{Results}
\begin{sloppypar}
We evaluated our method with two leading baselines: the original 3DGS~\citep{kerbl20233d} and DRGS~\citep{chung2024depth}. 
Our evaluation focused on both 2D novel view synthesis quality and 3D geometry reconstruction accuracy.

\textbf{Novel View Synthesis Quality.} 
We assessed the effectiveness of our method in \gls{nvs} through both qualitative and quantitative analyses, with particular attention to early-stage performance. 
Our method demonstrated consistent advantages throughout the first half of training iterations, and we showcase the results at iteration 9,000 (30\% of total iterations) as a representative example. 
Figure~\ref{fig:ignatius_comparison} provides detailed visual comparisons on two representative scenes (Ignatius and Truck) at this stage. 
The comparison presented both synthesized RGB images and their corresponding depth maps to demonstrate the overall reconstruction quality. 
Our method exhibited notably better geometric detail preservation during these early stages, especially evident in the highlighted regions (yellow boxes), and suggested more efficient optimization and faster convergence to high-quality reconstructions.

For quantitative evaluation, Table~\ref{tab:nvs_metrics} presents 2D evaluation metrics across different scenes at iteration 9,000. 
At this early stage, compared to the original \gls{3dgs}, our method showed improvements with PSNR gains of up to 2.815 dB in Barn scene and SSIM improvements of up to 0.050 in Caterpillar scene. 
Besides, our method demonstrated competitive results with DRGS in terms of average SSIM (0.600) and LPIPS (0.459). 

\begin{table*}[ht!]
    \centering
    \caption{Quantitative comparison of novel view synthesis quality on the \gls{tnt} dataset. The metrics include PSNR (dB), SSIM, and LPIPS. $\uparrow$ indicates higher is better, $\downarrow$ indicates lower is better. Best results are highlighted in \colorbox{DarkGreen!50}{\textbf{green}} and shown in \textbf{bold}.}
    \label{tab:nvs_metrics}
    \begin{tabular}{l|ccc|ccc|ccc}
        \toprule
        \multirow{2}{*}{Scene} & \multicolumn{3}{c|}{3DGS} & \multicolumn{3}{c|}{DRGS} & \multicolumn{3}{c}{Ours} \\
        \cmidrule{2-10}
        & PSNR$\uparrow$ & SSIM$\uparrow$ & LPIPS$\downarrow$ & PSNR$\uparrow$ & SSIM$\uparrow$ & LPIPS$\downarrow$ & PSNR$\uparrow$ & SSIM$\uparrow$ & LPIPS$\downarrow$ \\
        \midrule
        Barn & 17.212 & 0.588 & 0.545 & 19.987 & 0.633 & 0.463 & \cellcolor{DarkGreen!50}\textbf{20.027} & \cellcolor{DarkGreen!50}\textbf{0.638} & \cellcolor{DarkGreen!50}\textbf{0.461} \\
        Caterpillar & 16.061 & 0.587 & 0.558 & 18.698 & 0.553 & 0.456 & \cellcolor{DarkGreen!50}\textbf{18.836} & \cellcolor{DarkGreen!50}\textbf{0.567} & \cellcolor{DarkGreen!50}\textbf{0.452} \\
        Truck & 13.933 & 0.546 & 0.587 & \cellcolor{DarkGreen!50}\textbf{15.291} & 0.574 & \cellcolor{DarkGreen!50}\textbf{0.492} & 15.132 & \cellcolor{DarkGreen!50}\textbf{0.578} & 0.495 \\
        Ignatius & 16.142 & 0.452 & 0.537 & \cellcolor{DarkGreen!50}\textbf{17.918} & 0.502 & 0.466 & 17.813 & \cellcolor{DarkGreen!50}\textbf{0.510} & \cellcolor{DarkGreen!50}\textbf{0.452} \\
        Meeting Room & 16.938 & 0.677 & 0.508 & \cellcolor{DarkGreen!50}\textbf{19.746} & \cellcolor{DarkGreen!50}\textbf{0.727} & \cellcolor{DarkGreen!50}\textbf{0.401} & 18.058 & 0.698 & 0.417 \\
        Church & 15.619 & 0.568 & 0.555 & \cellcolor{DarkGreen!50}\textbf{17.098} & \cellcolor{DarkGreen!50}\textbf{0.612} & 0.479 & 16.916 & 0.609 & \cellcolor{DarkGreen!50}\textbf{0.474} \\
        \midrule
        Average & 15.984 & 0.570 & 0.548 & \cellcolor{DarkGreen!80}\textbf{18.123} & \cellcolor{DarkGreen!80}\textbf{0.600} & 0.460 & 17.797 & \cellcolor{DarkGreen!80}\textbf{0.600} & \cellcolor{DarkGreen!80}\textbf{0.459} \\
        \bottomrule
    \end{tabular}
\end{table*}

\begin{figure*}[ht!]
    \centering
    \begin{minipage}[b]{0.3\textwidth}
        \centering
        \includegraphics[width=\textwidth]{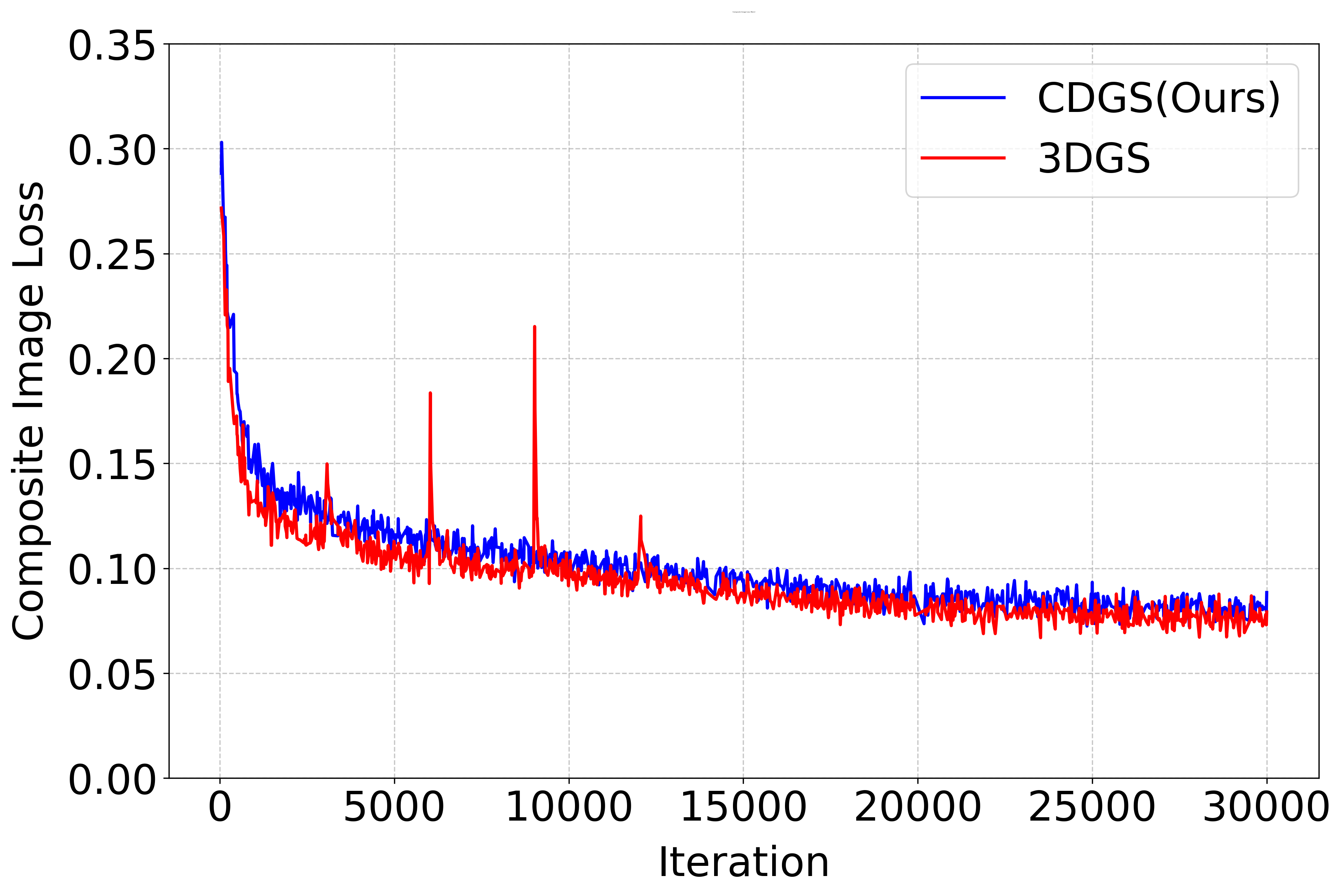}
        \subfigure{Barn}
    \end{minipage}%
    \begin{minipage}[b]{0.3\textwidth}
        \centering
        \includegraphics[width=\textwidth]{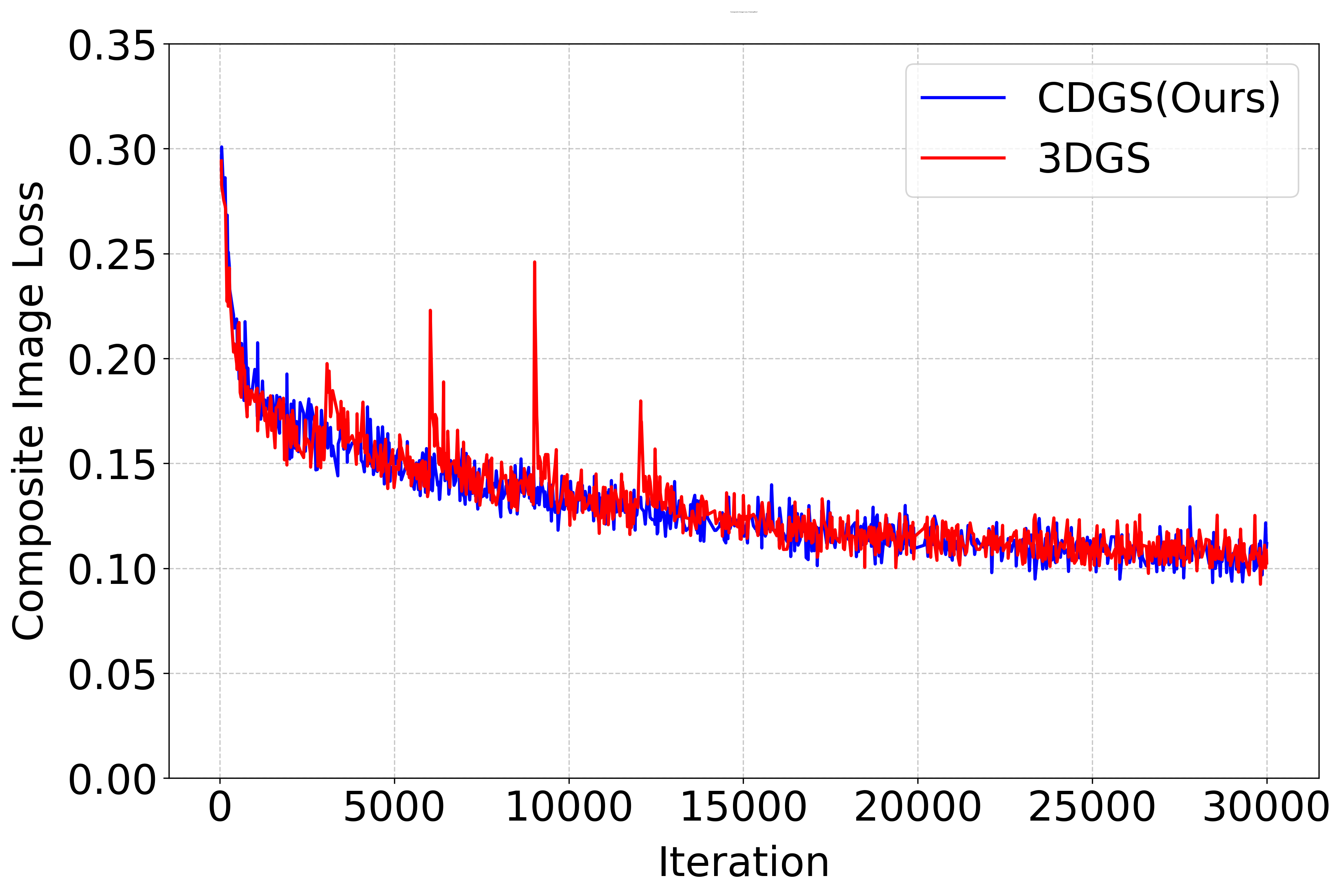}
        \subfigure{Caterpillar}
    \end{minipage}%
    \begin{minipage}[b]{0.3\textwidth}
        \centering
        \includegraphics[width=\textwidth]{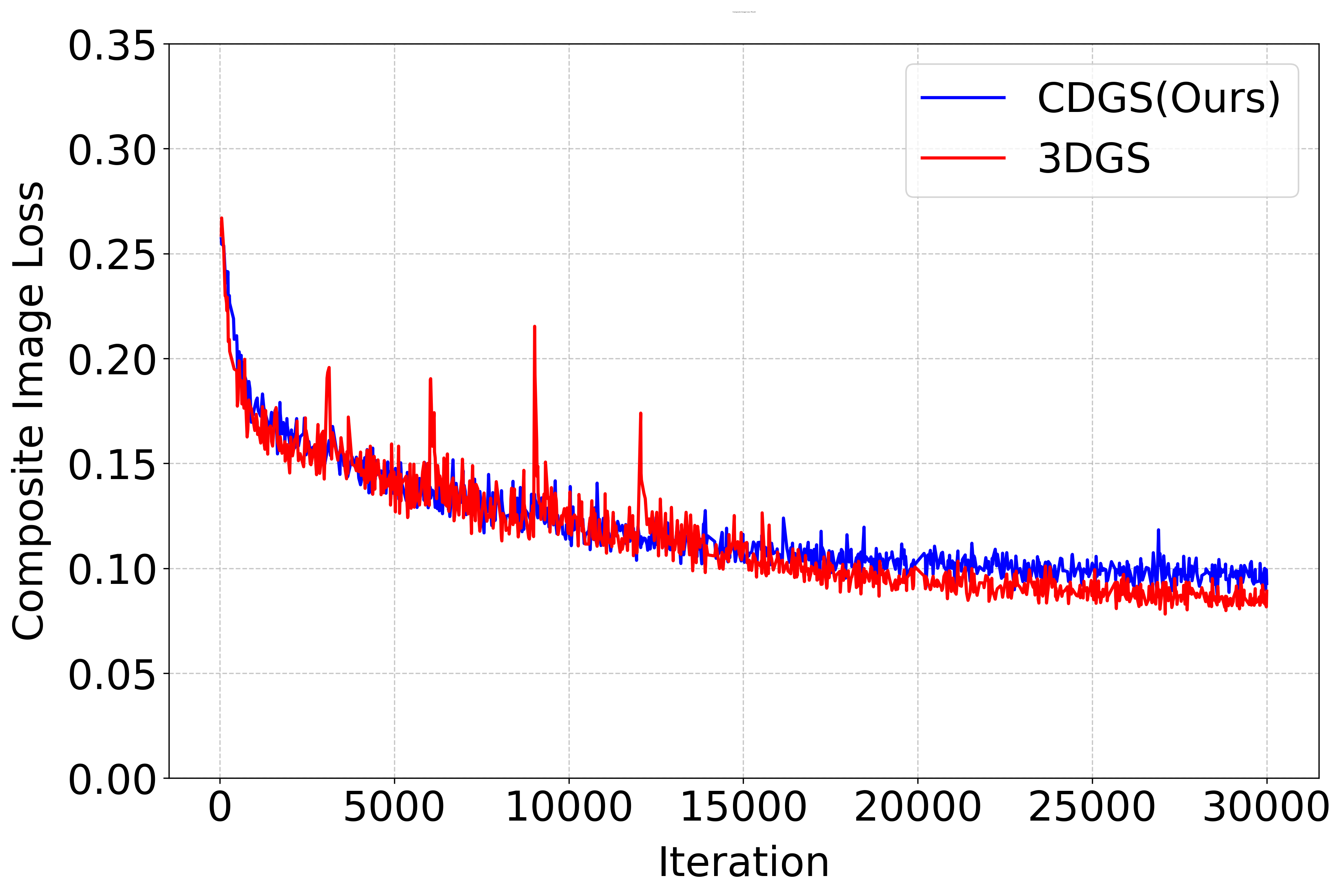}
        \subfigure{Truck}
    \end{minipage}
    
    \begin{minipage}[b]{0.3\textwidth}
        \centering
        \includegraphics[width=\textwidth]{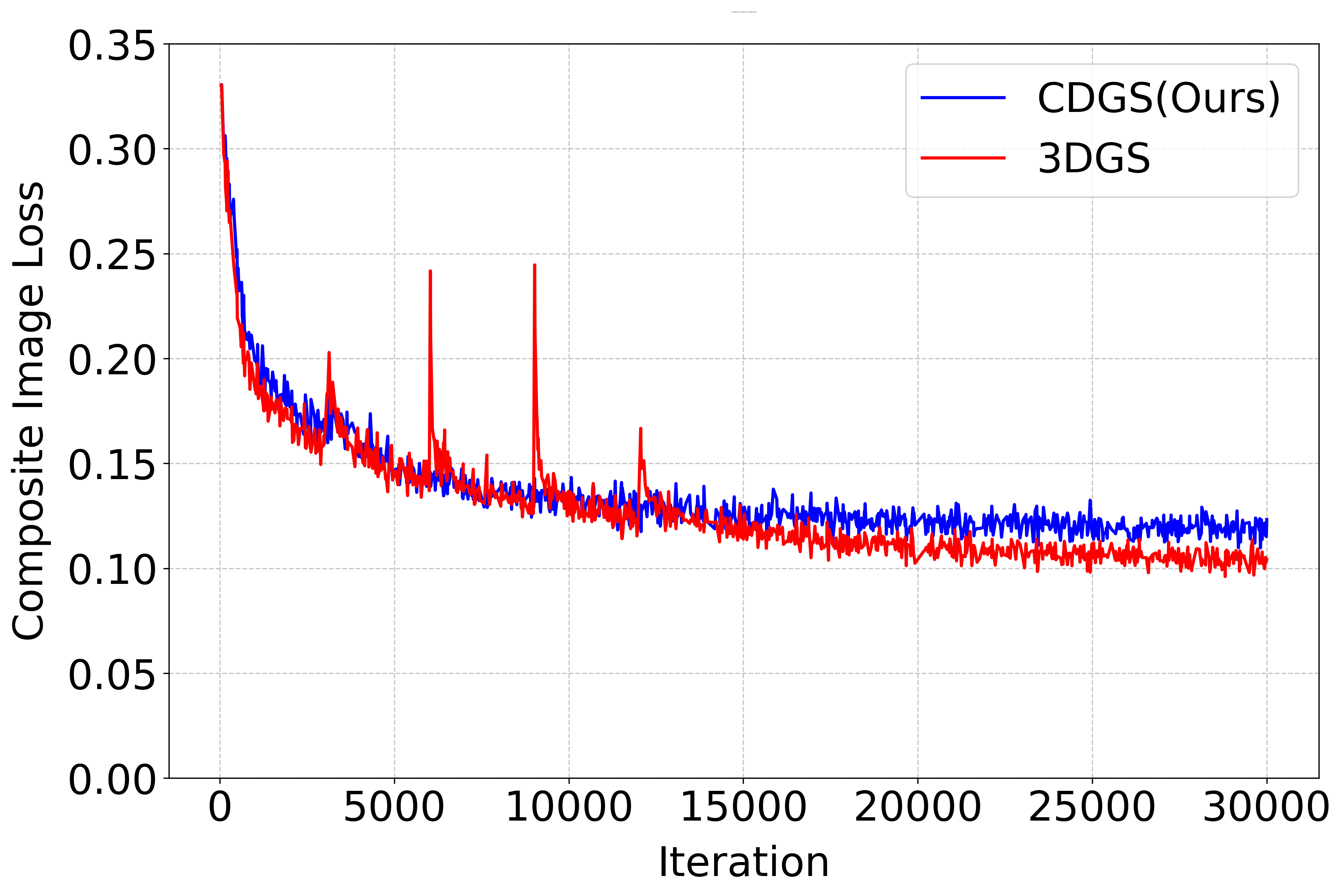}
        \subfigure{Ignatius}
    \end{minipage}
    \begin{minipage}[b]{0.3\textwidth}
        \centering
        \includegraphics[width=\textwidth]{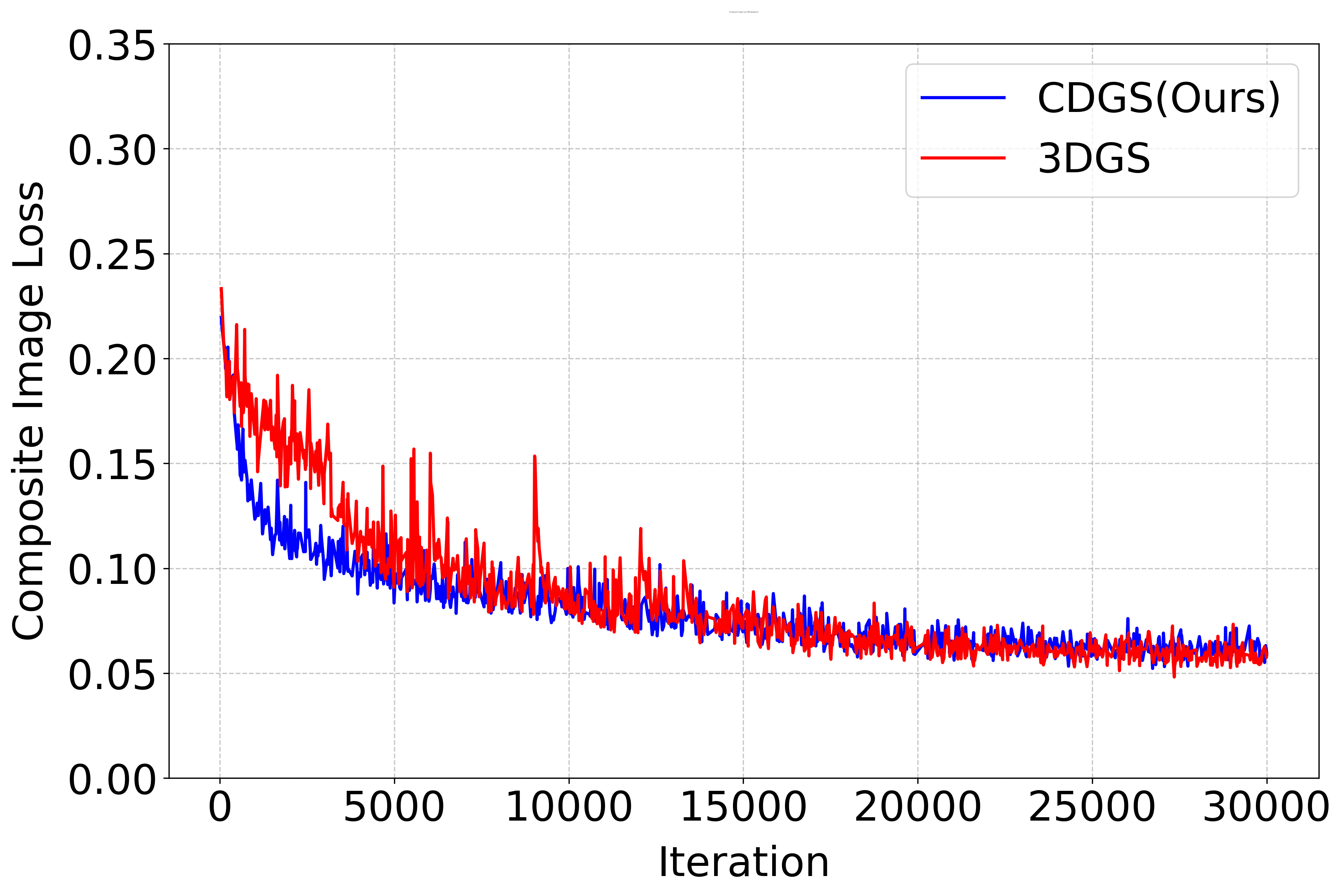}
        \subfigure{Meetingroom}
    \end{minipage}
    \begin{minipage}[b]{0.3\textwidth}
        \centering
        \includegraphics[width=\textwidth]{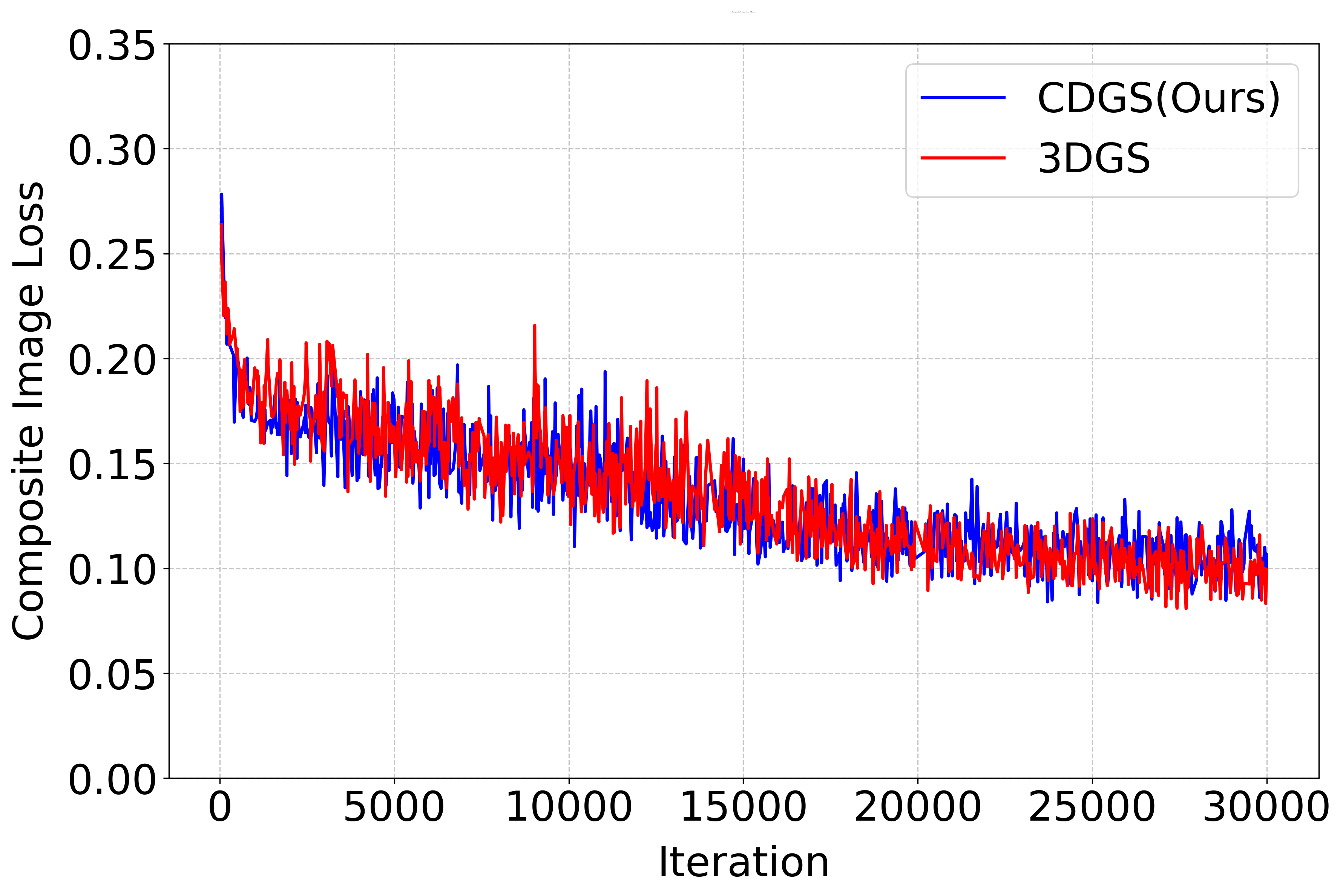}
        \subfigure{Church}
    \end{minipage}
    \caption{Comparison of composite image loss during training across different scenes from the \gls{tnt} dataset. Each plot shows the convergence behavior over 30,000 iterations, with our method (blue) and \gls{3dgs} (red). Lower values indicate better performance.}
    \label{fig:all_scenes_loss}
\end{figure*}

\textbf{Training Stability and Convergence.} 
Figure~\ref{fig:all_scenes_loss} illustrates the convergence behavior of the composite image loss during training across different scenes. 
The composite image loss combines L1 reconstruction loss and SSIM-based perceptual loss, providing a comprehensive measure of image synthesis quality. 
Our approach demonstrated more stable convergence patterns compared to the original \gls{3dgs}, with fewer fluctuations across all tested scenes.

\begin{figure*}[ht!]
    \centering
    \begin{minipage}[b]{0.3\textwidth}
        \centering
        \includegraphics[width=\textwidth]{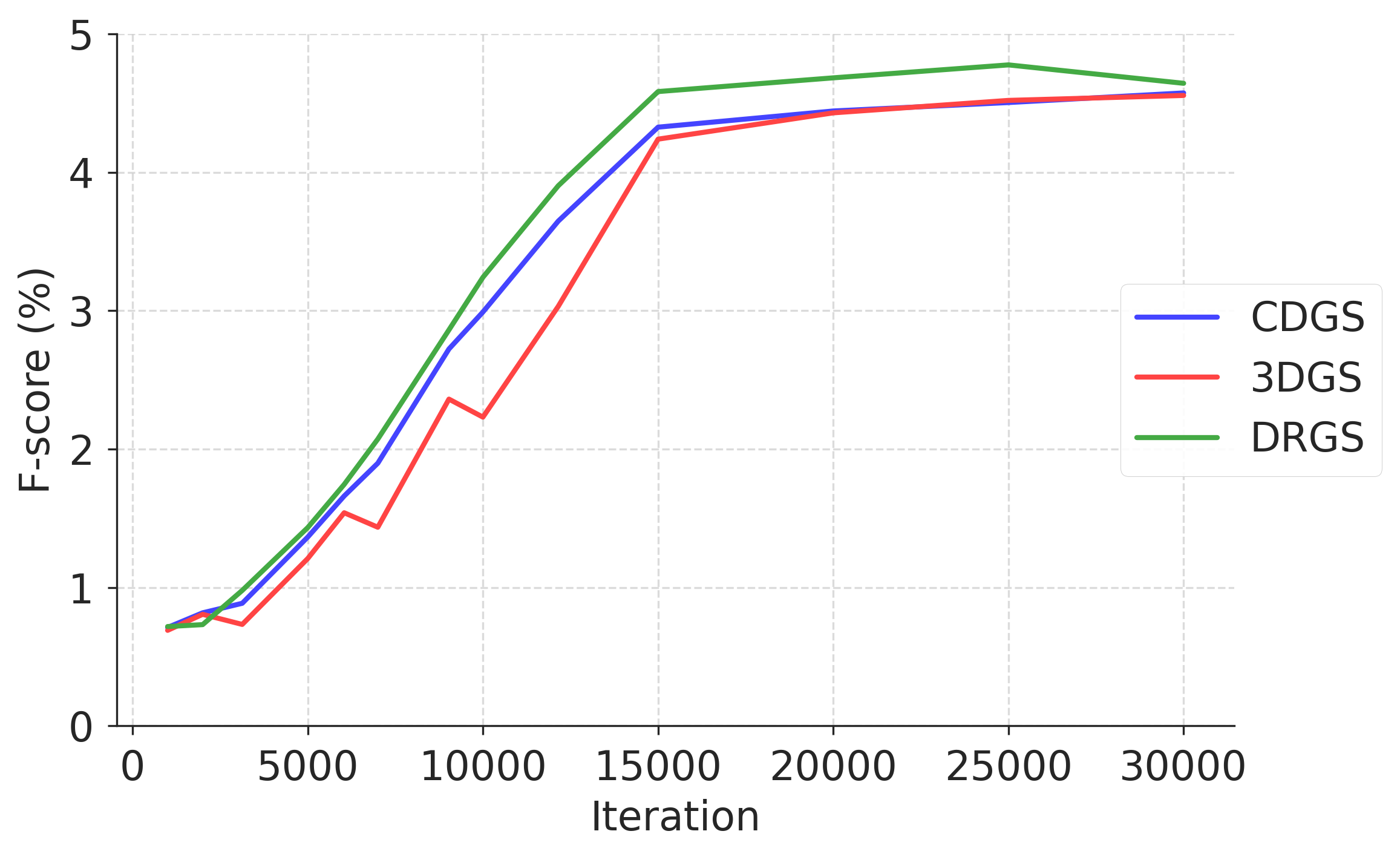}
        \subfigure{Barn}
    \end{minipage}%
    \begin{minipage}[b]{0.3\textwidth}
        \centering
        \includegraphics[width=\textwidth]{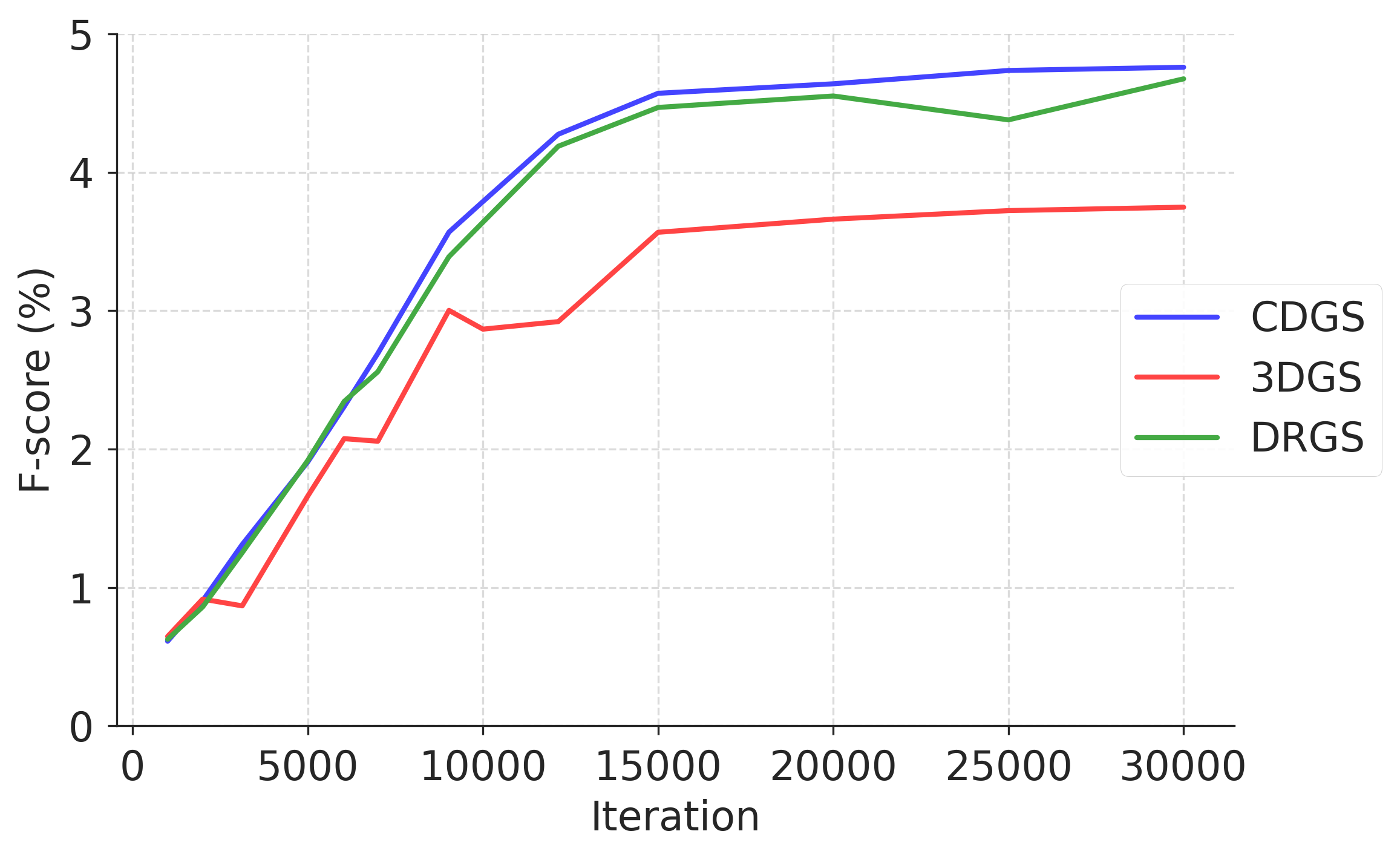}
        \subfigure{Caterpillar}
    \end{minipage}%
    \begin{minipage}[b]{0.3\textwidth}
        \centering
        \includegraphics[width=\textwidth]{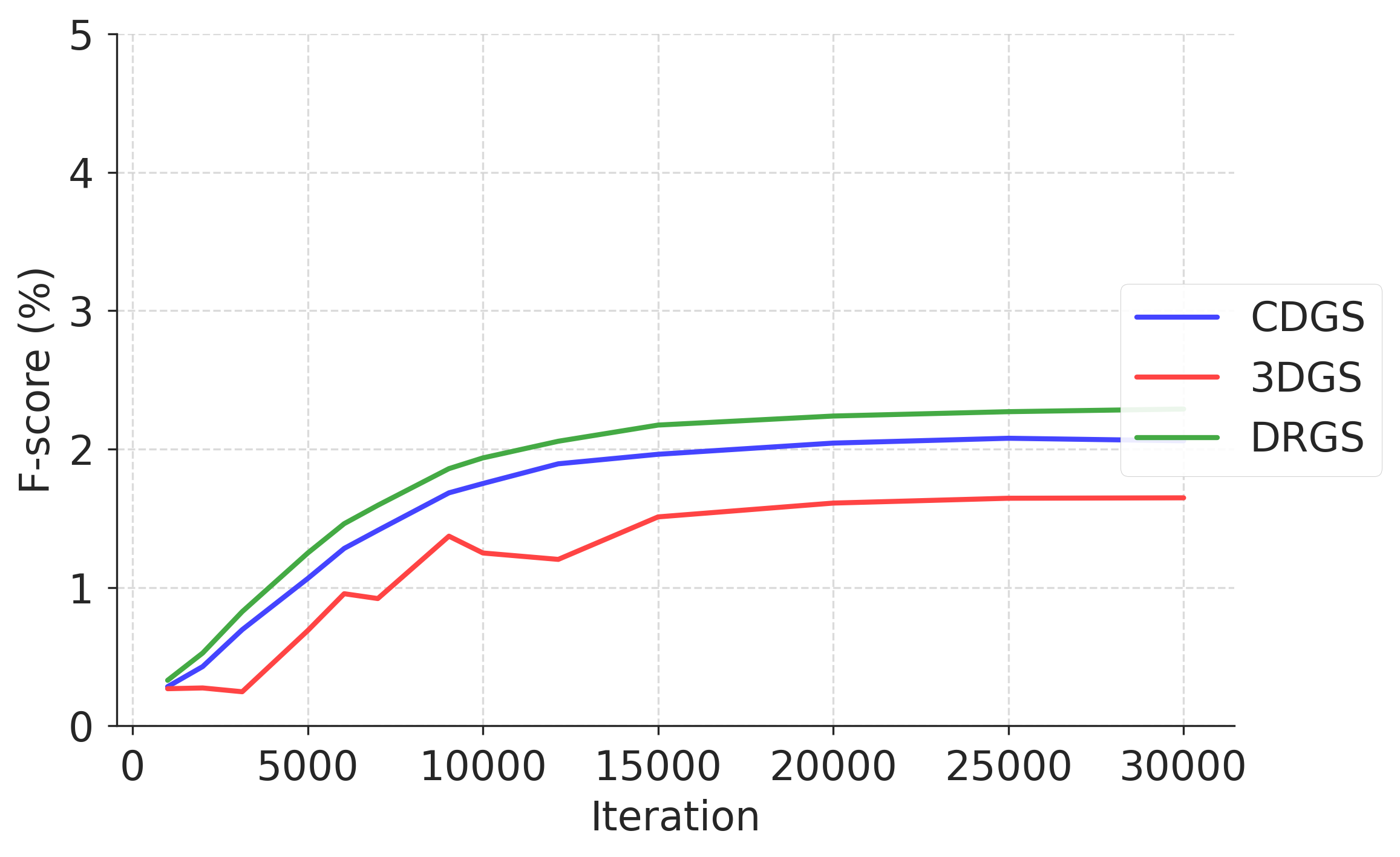}
        \subfigure{Meetingroom}
    \end{minipage}
    \caption{F-score evolution during training for Barn, Caterpillar, and Meeting Room scenes from the TnT dataset. Our method (CDGS, in blue), DRGS (in green), and 3DGS (in red). The y-axis range is set to 0-5\%.}
    \label{fig:all_scenes_fscore}
\end{figure*}

\textbf{Geometric Accuracy.} 
We evaluated the geometric accuracy through three complementary metrics: F-score evolution, geometric completeness, and M3C2 distance analysis. 
Figure~\ref{fig:all_scenes_fscore} illustrates the F-score evolution during training for three representative scenes (Barn, Caterpillar, and Meeting Room). 
The results showed that both depth-aware methods (DRGS and CDGS) achieved higher F-scores compared to the original \gls{3dgs} across these scenes. 

For a comprehensive evaluation at iteration 15,000 (where most scenes demonstrate stable performance), Table~\ref{tab:geometric_completeness} presents the geometric completeness results. 
While the original 3DGS achieves higher precision values across all scenes (with the highest being 0.248 in Ignatius), both depth-aware methods show better recall and F-score performance.
\begin{table*}[ht!]
    \centering
    \caption{Geometric completeness comparison at iteration 15,000. F-score values are in percentage (\%). Best results are shown in \textbf{bold}.}
    \label{tab:geometric_completeness}
    \setlength{\tabcolsep}{5pt}
    \begin{tabular}{lcccclccccclcccc}
        \toprule
        \multirow{2}{*}{Scene} & \multicolumn{3}{c}{3DGS} & \phantom{a} & \multicolumn{3}{c}{DRGS} & \phantom{a} & \multicolumn{3}{c}{CDGS (Ours)} \\
        \cmidrule(lr){2-4} \cmidrule(lr){6-8} \cmidrule(lr){10-12}
        & Precision & Recall & F-score(\%) & & Precision & Recall & F-score(\%) & & Precision & Recall & F-score(\%) \\
        \midrule
        Barn & \textbf{0.143} & 0.025 & 4.2 & & 0.105 & \textbf{0.029} & \textbf{4.5} & & 0.099 & 0.028 & 4.4 \\
        Caterpillar & \textbf{0.113} & 0.021 & 3.5 & & 0.102 & 0.028 & 4.4 & & 0.106 & \textbf{0.029} & \textbf{4.5} \\
        Truck & \textbf{0.039} & 0.006 & 1.1 & & 0.031 & 0.006 & 1.0 & & 0.033 & \textbf{0.007} & \textbf{1.2} \\
        Ignatius & \textbf{0.248} & \textbf{0.072} & \textbf{11.3} & & 0.160 & 0.033 & 5.4 & & 0.166 & 0.033 & 5.4 \\
        Meetingroom & \textbf{0.078} & 0.008 & 1.5 & & 0.067 & \textbf{0.013} & \textbf{2.2} & & 0.067 & 0.011 & 1.9 \\
        Church & \textbf{0.061} & 0.006 & 1.1 & & 0.052 & \textbf{0.011} & 1.9 & & 0.054 & \textbf{0.013} & \textbf{2.1} \\
        \midrule
        Average & \textbf{0.114} & 0.023 & \textbf{3.8} & & 0.086 & \textbf{0.020} & 3.2 & & 0.088 & \textbf{0.020} & 3.3 \\
        \bottomrule
    \end{tabular}
\end{table*}

\begin{figure*}[ht!]
    \centering
    \begin{minipage}[b]{0.3\textwidth}
        \centering
        \includegraphics[width=\textwidth]{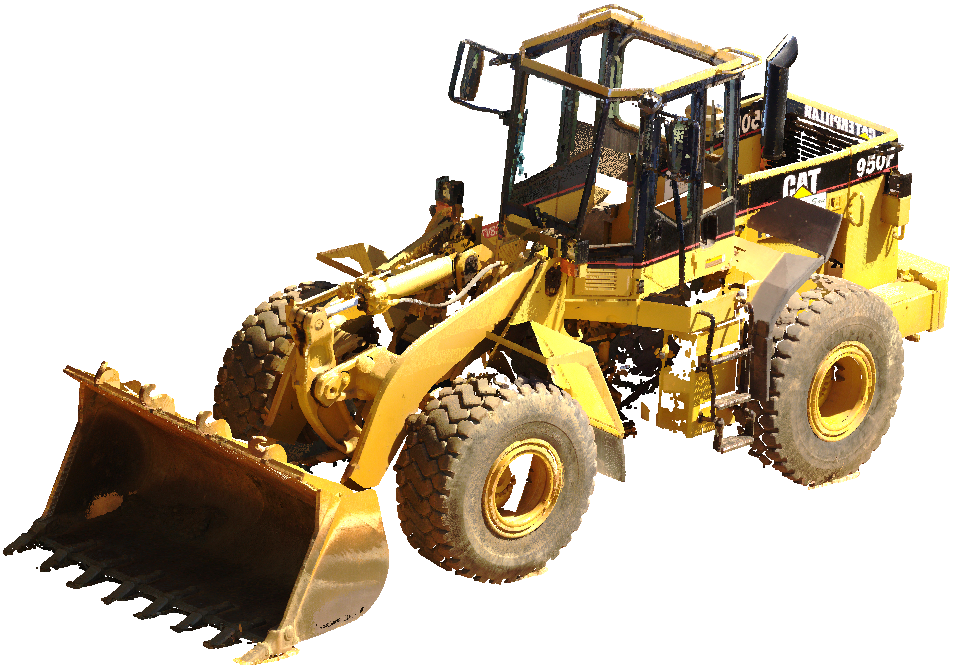}
        \subfigure{Ground truth point cloud}
    \end{minipage}%
    \begin{minipage}[b]{0.3\textwidth}
        \centering
        \begin{tikzpicture}
            \node[anchor=west] at (0,0) {
                \includegraphics[width=0.9\textwidth]{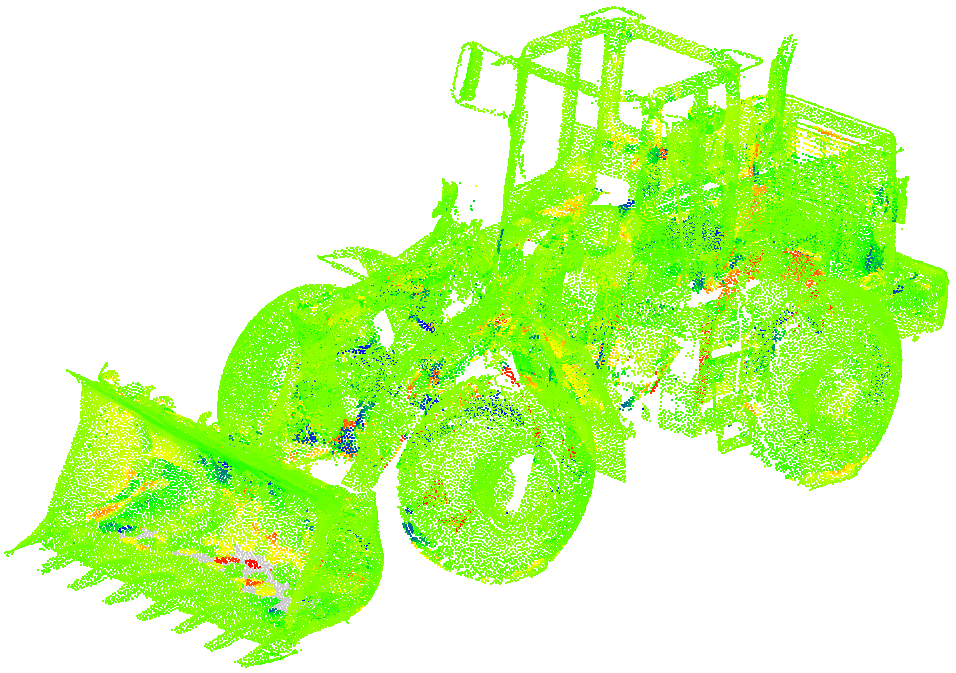}
            };
            \node[anchor=west] at (0.95\textwidth,0) {
                \begin{tikzpicture}[scale=0.5]
                    \begin{axis}[
                        hide axis,
                        scale only axis,
                        height=1.0\textwidth,
                        width=0.1\textwidth,
                        colormap={custom}{
                            color(0cm)=(blue);
                            color(2.5cm)=(green);
                            color(5cm)=(red)
                        },
                        colorbar right,
                        point meta min=-0.4,
                        point meta max=0.4,
                        colorbar style={
                            title=Distance (m),
                            ytick={-0.4,-0.2,0,0.2,0.4},
                            title style={yshift=-1ex},
                        }
                    ]
                    \addplot[draw=none] coordinates {(0,0)};
                    \end{axis}
                \end{tikzpicture}
            };
        \end{tikzpicture}
        \subfigure{M3C2 distance visualization}
    \end{minipage}
    \caption{M3C2 distance analysis visualization of our CDGS reconstruction on the Caterpillar scene. The color bar indicates the signed distances to the ground truth surface, where green represents small distances ($\pm$0.05 m), blue indicates negative deviations \\ (up to -0.4 m), and red shows positive deviations (up to 0.4 m).}
    \label{fig:m3c2_vis}
\end{figure*}

Following \citet{jager2024hologs}, we further employed the M3C2 metric to evaluate point cloud accuracy (Table~\ref{tab:m3c2_distance}). 
Our method achieved the lowest RMSE values in five out of six scenes, with an average RMSE of 0.120 meters. 
Figure~\ref{fig:m3c2_vis} provides a qualitative visualization of the M3C2 distance analysis on the Caterpillar scene. 
The color-coded visualization shows the signed distances between our reconstructed point cloud and the ground truth, where green indicates small distances, blue represents inward deviations, and red shows outward deviations. 
The predominant green coloring in our reconstruction demonstrates the high geometric fidelity achieved by our method, particularly in preserving the complex mechanical structures of the scene.

\begin{table}[ht!] 
    \centering
    \caption{Point cloud distance analysis using M3C2 metric. RMSE values are in meters. Lower values indicate better accuracy. Best results are shown in \textbf{bold}.}
    \label{tab:m3c2_distance}
    \setlength{\tabcolsep}{6pt}
    \begin{tabular}{lccc}
        \toprule
        Scene & 3DGS & DRGS & CDGS (Ours) \\
        \midrule
        Barn & 0.084 & 0.077 & \textbf{0.076} \\
        Caterpillar & 0.082 & 0.081 & \textbf{0.080} \\
        Truck & 0.201 & 0.199 & \textbf{0.195} \\
        Ignatius & \textbf{0.032} & 0.035 & 0.035 \\
        Meetingroom & 0.158 & 0.159 & \textbf{0.157} \\
        Church & 0.187 & 0.178 & \textbf{0.177} \\
        \midrule
        Average & 0.124 & 0.122 & \textbf{0.120} \\
        \bottomrule
    \end{tabular}
\end{table}

\textbf{Ablation Studies.} 
We conducted ablation studies on the Ignatius scene to evaluate the contribution of each key component in our method (Table~\ref{tab:ablation}). 
Removing the confidence map mechanism led to performance degradation across all metrics at 30,000 iterations, with PSNR decreasing from 18.22 dB to 18.04 dB. 
Similarly, replacing our adaptive depth loss weighting scheme with fixed scale factors resulted in more substantial decreases, with PSNR dropping by 1.12 dB (from 18.22 dB to 17.10 dB) and SSIM declining from 0.519 to 0.484. 
At 15,000 iterations, our full method achieved a PSNR of 18.04 dB and LPIPS of 0.411, compared to 17.80 dB/0.424 without confidence map and 17.10 dB/0.458 with fixed depth loss. 

\begin{table}[ht!]
    \centering
    \caption{Ablation study results on the Ignatius scene at 30,000 iterations. Best results are shown in \textbf{bold}.}
    \label{tab:ablation}
    \begin{tabular}{lccc}
        \toprule
        Method & PSNR$\uparrow$ & SSIM$\uparrow$ & LPIPS$\downarrow$ \\
        \midrule
        Ours (Full) & \textbf{18.225} & \textbf{0.519} & \textbf{0.384} \\
        w/o Confidence Map & 18.040 & 0.509 & 0.396 \\
        w/ Fixed Depth Loss & 17.100 & 0.484 & 0.437 \\
        \bottomrule
    \end{tabular}
\end{table}
\end{sloppypar}
\subsection{Discussion}
\begin{sloppypar}
The experimental results demonstrate the effectiveness of our method from multiple perspectives. 
For 2D novel view synthesis, our confidence-aware depth regularization enables better geometric detail preservation during early training stages (30\% of total iterations) by effectively identifying and emphasizing reliable depth regions. 
While DRGS achieves slightly better average PSNR (18.12 dB vs. our 17.80 dB), our method shows particularly strong performance in scenes with complex geometric structures, such as the Barn (20.03 dB) and Caterpillar (18.84 dB) scenes, suggesting the effectiveness of our confidence-guided approach in handling challenging scenarios.

The geometric evaluation reflects a trade-off between precision and recall. 
The original 3DGS tends to generate more conservative but accurate reconstructions, achieving the highest precision across all scenes (e.g., 0.248 for Ignatius and 0.143 for Barn), while depth-aware methods achieve better geometric completeness through improved recall (e.g., increasing from 0.025 to 0.028 in Barn scene). 
This trade-off underscores the role of depth supervision in recovering more complete geometric structures. This is further supported by our M3C2 analysis, which shows that our method achieves the lowest RMSE values in five out of six scenes. On average, our approach achieves an RMSE of 0.120 meters, outperforming both 3DGS (0.124 meters) and DRGS (0.122 meters). 

The convergence analysis demonstrates the stability benefits of depth-aware optimization. 
Both DRGS and our method achieve smoother convergence compared to the original \gls{3dgs}, with particularly stable behavior during 15-50\% of total training iterations. 
This enhanced stability, combined with faster geometric convergence shown in F-score evolution (reaching F-scores of 4.4\% for Barn and 4.5\% for Caterpillar at iteration 15,000), makes our method particularly suitable for applications with training time constraints. 

The ablation studies highlight two key aspects of our method. 
First, the confidence map mechanism, while yielding a modest improvement in final quality (PSNR increased from 18.04 dB to 18.22 dB), plays a critical role in establishing reliable geometric structures during early training. 
Second, replacing the adaptive depth loss weighting scheme with fixed weights significantly degrades performance, with PSNR dropping from 18.22 dB to 17.10 dB and SSIM decreasing from 0.519 to 0.484, emphasizing the importance of dynamic supervision. 

Despite these improvements, our method has limitations. Performance varies across scenes, particularly in indoor environments with complex lighting (e.g., the Meeting Room scene shows a 1.69 dB lower PSNR than DRGS). 
Additionally, reconstructing fine geometric details remains challenging in occluded areas with limited depth information. 
These limitations suggest future research directions, including refined confidence estimation for challenging scenarios and exploration of additional geometric priors for improved reconstruction accuracy.
\end{sloppypar}
\section{Conclusions}\label{Conclusions}
\begin{sloppypar}
In this work, we have presented CDGS, a confidence-aware depth regularization framework for enhancing \gls{3dgs}. 
Through the integration of aligned monocular depth estimates and multi-feature confidence assessment, our method achieves more stable optimization and improved geometric reconstruction. 
The experimental results demonstrate that our approach effectively preserves geometric details during early training stages and maintains competitive performance in both 2D synthesis and 3D reconstruction tasks. 
We assume this work will encourage further research on leveraging depth information and geometric constraints to enhance the 3D reconstruction accuracy of Gaussian Splatting methods. 
Moreover, we expect this work will facilitate the development of efficient and accurate 3D reconstruction systems for real-world applications such as digital twin creation, heritage preservation, or forestry applications. 

The ablation studies confirm the importance of confidence-aware depth supervision and adaptive loss weighting in achieving better optimization behavior. 
While showing promising results, our method still faces challenges in certain scenarios, particularly in indoor environments with complex lighting conditions and regions with intricate geometric structures. 
Future work will focus on advancing confidence estimation techniques to handle challenging scenes and exploring additional geometric priors to further improve reconstruction accuracy. 
\end{sloppypar}
{
	\begin{spacing}{1.17}
		\normalsize
		\bibliography{ISPRSguidelines_authors} 
	\end{spacing}
}

\end{document}